\documentclass[%
 reprint,
 amsmath,amssymb,
 aps,
pre,
floatfix
]{revtex4-2}
\usepackage{color}
\usepackage{graphicx}% Include figure files
\usepackage{dcolumn}% Align table columns on decimal point
\usepackage{bm}% bold math
\begin{document}

\title{An Application of Quantum Machine Learning on Quantum Correlated Systems: Quantum Convolutional Neural Network as a Classifier for Many-Body Wavefunctions from the Quantum Variational Eigensolver }% Force line breaks with \\

\author{Nathaniel Wrobel}%
\affiliation{Department of Physics \& Astronomy, Louisiana State University, Baton Rouge, Louisiana 70803, USA}%

\author{Anshumitra Baul}%
\affiliation{Department of Physics \& Astronomy, Louisiana State University, Baton Rouge, Louisiana 70803, USA}%
%\affiliation{Center for Computation & Technology, Louisiana State University, Baton Rouge, LA 70803, USA}

\author{Juana Moreno}
\affiliation{Department of Physics \& Astronomy, Louisiana State University, Baton Rouge, Louisiana 70803, USA}%
\affiliation{Center for Computation and Technology, Louisiana State University, Baton Rouge, LA 70803, USA}

\author{Ka-Ming Tam}
\affiliation{Department of Physics \& Astronomy, Louisiana State University, Baton Rouge, Louisiana 70803, USA}%
\affiliation{Center for Computation and Technology, Louisiana State University, Baton Rouge, LA 70803, USA}

%\thanks{Selected Paper prepared for presentation at the 201X Agricultural \& Applied Economics Association Annual Meeting}}

\date{\today}% It is always \today, today,

\begin{abstract}

Machine learning has been applied on a wide variety of models, from classical statistical mechanics to quantum strongly correlated systems for the identification of phase transitions. The recently proposed quantum convolutional neural network (QCNN) provides a new framework for using quantum circuits instead of classical neural networks as the backbone of classification methods. We present here the results from training the QCNN by the wavefunctions of the variational quantum eigensolver for the one-dimensional transverse field Ising model (TFIM). We demonstrate that the QCNN identifies wavefunctions which correspond to the paramagnetic phase and the ferromagnetic phase of the TFIM with good accuracy. The QCNN can be trained to predict the corresponding ‘phase’ of wavefunctions around the putative quantum critical point, even though it is trained by wavefunctions far away from it. This provides a basis for exploiting the QCNN to identify the quantum critical point.

\end{abstract}

\maketitle


%\tableofcontents

% %%%%%%%%%%%%%%%%%%%%%%%%%%%%%%%%%%%%%%%%%%%%%%%%%%%%%%%%%%
% %%%%%%%%%%%%%%%%%%%%%%%%%%%%%%%%%%%%%%%%%%%%%%%%%%%%%%%%%%
% BODY OF THE DOCUMENT
% %%%%%%%%%%%%%%%%%%%%%%%%%%%%%%%%%%%%%%%%%%%%%%%%%%%%%%%%%%
% %%%%%%%%%%%%%%%%%%%%%%%%%%%%%%%%%%%%%%%%%%%%%%%%%%%%%%%%%%
\section{Introduction}
Machine learning (ML) and quantum computing (QC) are among the most notable topics which may have significant impacts on many different fields of physics. ML has grown into a powerful tool for scientific and academic use in the age of big data. The progress of QC, in particular the realization of quantum computers with tens of qubits, may grant a new opportunity for the study of challenging problems in strongly correlated many-body physics among other applications.

ML has been applied to a wide range of topics in physics and other branches of science and engineering. It has seen an explosive growth of diverse applications in the past decade or so. This is principally driven by the availability of a large data set and accessible libraries for sophisticated deep learning methods based on neural networks \cite{keras,tf}. Among different types of neural networks, the convolution neural network (CNN) is widely used \cite{Lecun_etal_1998}. Unlike the conventional dense or fully connected neural network, CNN emphasises the information of local correlation. It serves as a high performance classifier for computer vision. Image identification is a main topic of classifiers. Most images have a certain level of spatial correlation. The CNN is designed to utilize those local spatial correlations in the input data. In practice, many physical data also possess strong spatial correlation, therefore it is in hindsight not surprising that the CNN has seen many successful applications in physics.

In ML, particularly, CNN has been adopted to the identification of  phase transitions of classical statistical models from snapshots of classical Monte Carlo or molecular dynamics configurations, as well as configurations from quantum Monte Carlo of strongly correlated systems \cite{ising_vae,ising_vae_2,ising_vae_3,Wang_2016,Walker_etal_2018}. 
%Similarly, quantum Monte Carlo configurations can also be used to trained the CNN \cite{Chng_etal_2017}. 
Attempts have also been made for using the quantum wavefunction from exact diagonalization \cite{Hsu_etal_2018,Schindler_etal_2017,Zhang_Wang_Wang_2019,Walker_Tam_2021,Carrasquilla_etal_2017,Munoz_Bauza_etal_2020,Shiina_etal_2020,Broecker_etal_2017,Lozano-Gomez_etal_2020,Torlai_Melko_2016,Morningstar_Melko_2017,Alexandrou_etal_2020,Wetzel_etal_2017}. Recent studies further involve in feeding spatially resolved experimental data such as those from scanning tunneling microscopy for identifying different phases of materials \cite{Zhang_etal_2019}.

While a quantum computer for fault tolerant quantum computations which can supersede the best classical computer for many tasks may not be available in a near future, noisy quantum computers with tens of qubits are immediately available. Noisy intermediate scale quantum (NISQ) computers are also likely to be feasible in the near future \cite{Preskill_2018}. They open up a new opportunity for using quantum computation for solving problems in a strikingly different fashion than classical numerical simulations. Among the various methods which are feasible in such NISQ computers, the variational quantum eigensolver \cite{Peruzzo_etal_2014} and the general idea of quantum approximation optimization algorithm are among the most promising proposals \cite{Farhi_etal_2014}. 

Enormous amounts of effort for addressing problems in optimization, chemistry, and even strongly correlated systems have started to rise in the past few years \cite{Fedorov_etal_2021,McClean_etal_2016}. Conceptually, the approach is based on a quantum state with parameters. The quantum computer is used to calculate the expectation value of a given quantum state to the quantity produced for optimization \cite{McClean_etal_2016}. This can be a cost function in general optimization problems, or the ground state energy of a molecule. In general, calculating such expectation value scales exponentially with respect to the problem size by classical methods, the quantum computer offers an opportunity to speedup such calculation. The parameter is then optimized by a classical optimization algorithm. The idea of variational methods are not limited to only ground state calculation, it is a general concept which can be used to mimic any operator in a variational sense. For example, the quantum dynamics based on solving the Schrodinger equation can be estimated by the variational method \cite{Yuan_etal_2019,McArdle_etal_2019}.

Variational methods have been widely adopted in condensed matter physics. Specifically, the variational quantum Monte Carlo (VMC) is one of the major numerical methods for solving correlated systems \cite{Feynman_1955,Ceperley_etal_1977,Casula_etal_2004,Ceperley_etal_1977,Umrigar_etal_1988,Yokohama_etal_1987,Edegger_etal_2007}.
The quantum expectation values for the ground state energy are  calculated by Monte Carlo. The minimization of ground state energy with respect to the variational parameters in the wavefunction is proceeded by the multivariable minimization method.  Its main advantage is due to the absence of the minus sign problem which hinders most quantum Monte Carlo methods for fermion problems. 

The VQE provides a new framework for sidestepping the computational intensive part of the conventional VMC method in calculating quantum expectation values by quantum computers \cite{McClean_etal_2016}. The wavefunctions represented in quantum circuits also provide new opportunities as well as new challenges due to the different nature of the wavefunctions that have been used in the conventional VMC \cite{Lee_etal_2018}. It is worthwhile to remark that most of the numerical methods for finding the ‘ground state’ of a many-body system are based on a nonunitary propagation of a trial state, a typical example is the projection quantum Monte Carlo \cite{Ceperley_1995}. 

From the viewpoint of utilizing quantum computing approaches for strongly correlated systems, calculating the ground state energy alone is often not sufficient to reveal much detail of the system. A particular interesting problem is the possibility of quantum phase transitions at zero temperature by tuning the parameters in the Hamiltonian \cite{Sachdev_2011,Hertz_1976}. Ground state energy, in particular, of rather small system sites that could be simulated in the near future, does not provide a direct answer for determining a quantum phase transition. Constructing an order parameter corresponding to the known broken symmetry in the thermodynamic limit allows a direct access to phase transitions. Unfortunately, given the small system size and the nature of a second-order phase transition of quantum phase transitions, an order parameter alone is often a rather obscure way to tell whether the systems possess a phase transition. This challenge leads to the development of the finite size scaling method \cite{Fisher_Barber_1972}, however for rather small system sizes that one can simulate, a proper finite size scaling may not be feasible. Moreover, there are systems in which phase transitions do not show up in local order parameters or the order parameter is simply unknown \cite{Senthil_etal_2004,Aeppli_etal_2020}.

The notion of using a quantum circuit as a classifier or clustering algorithm started to draw attention about a decade ago \cite{Lloyd_etal_2013,Gambs_2008}. The concepts of classical neural network had been filtered into the idea of quantum classifiers in the past few years \cite{Baul_etal_2021,Huang_etal_2021,Cappelletti_etal_2020,Belis_etal_2021,Sen_etal_2021,Park_etal_2021,Blank_etal_2020,Schuld_etal_2020,Miyahara_etal_2021,Blance_etal_2021,Grant_etal_2018,LaRose_etal_2020,Du_etal_2021,Abohashima_etal_2020,Chen_etal_2020,Farhi_etal_2018}. The classical neural network consists of links and neuron units represented by activation functions, which are organized in a layered structure. The key properties of CNN are translational invariant convolution and pooling layers, each characterized by a constant number of parameters (independent of system size), and sequential data size reduction (i.e., a hierarchical structure) \cite{Cong_etal_2019}.

For the purpose of using a neural network as a classifier, typically the number of neuron units decreases at each layer, until one or a few units remains in the output layer. Thus, each layer can be regarded as a pooling layer in the sense that the number of inputs to the neuron units at each layer is smaller than the number of outputs. It is, in a sense, a method for compressing and reducing the degrees of freedom, thus the suggestion that the concept of renormalization group can be pertinent in certain neural networks. 

For most space-dependent data, there is a nonzero spatial correlation. Particularly short range correlations appear in data from images of objects to physical systems, like spin correlation and spatial correlation of the positions of atoms or molecules in a solid phase and even in a liquid phase. Therefore, it is perhaps not surprising that the CNN has seen many applications in physics in learning patterns from statistical models to strongly correlated systems. 

The CNN is realized by introducing a so-called convolutional layer in each layer of activation functions of a dense neural network. The purpose is to extract ‘hidden’ information by some combination of local data, which is missing in the standard dense neural network. In practice, the combination is a weighted sum of local data. 

A simple analogue can be drawn to the quantum circuit by replacing the links and activation by the quantum links and the quantum gates, respectively \cite{Cong_etal_2019}. The principal structure of a QCNN is composed of two distinct types of layers. First, the pooling layer, which reduces the degree of freedom, can be replaced by multi-qubit gates. The simplest possibility is the CNOT gate \cite{Cong_etal_2019}. Second, the convolution layer in the CNN can be replaced by the multi-qubit quantum gates among nearby qubits. Thus, the QCNN can be understood naively as a quantum neural network classifier with convolutional layers in which `convolution’ between nearby qubits can be processed.

In short, a quantum circuit model is introduced which extends the key properties of the classical CNN to the quantum domain. The circuit’s input is an quantum state. A convolution layer applies a single quasi-local unitary in a translationally invariant way for finite depth. A fraction of qubits are measured for pooling, and their outcomes determine unitary rotations which are applied to nearby qubits \cite{Cong_etal_2019}. Hence, the nonlinearities in QCNN arise from reducing the number of degrees of freedom. Convolution and pooling layers are performed until the system size is sufficiently small. And then a fully connected layer is applied as a unitary function on the remaining qubits if needed. The outcome of the circuit is finally obtained by measuring a fixed number of output qubits. Likewise, in the classical CNN, circuit structures (i.e., hyperparameters of QCNN) such as the number of convolution and pooling layers are fixed. 

Recent studies have shown that quantum enhanced machine learning is a promising approach for recognizing the phase of matter \cite{Uvarov_etal_2020}. An interesting question is whether the QCNN method can be used to identify different phases of a quantum many-body system. That is the first step towards applying it for detecting quantum phase transitions. Using ML to identify phases by inputting the wavefunction is a challenge, as the Hilbert space of the system increases exponentially with respect to the system size. A practical method to bypass such a challenge is to consider the reduced density matrix or some other derived quantities based on the wavefunction \cite{Schindler_etal_2017}. 

An evident advantage of the QCNN approach is that the input is naturally quantum mechanical, the wavefunction does not need to be written down as a classical vector, which dimension grows exponentially with respect to the system size, to be fed to the ML method. Whereas the disadvantage or perhaps just an unknown factor is that the wavefunction is not calculated exactly, there is no control parameter to systematically improve the wavefunction.

It has to be input as some form of quantum circuit, and the one which is most promising in the NISQ is the VQE. The main purpose of the present manuscript is to present a study of a many-body quantum system solved by VQE, and use the QCNN to identify the VQE wavefunction corresponding to the different phases of the model to provide a possible framework for extracting quantum critical points. 

This paper is organized as follows. In section II, we briefly describe the transverse field Ising model (TFIM). In Section III, the data from the VQE of the TFIM is discussed and the structure of the QCNN is presented. The results from the variational autoencoder are described in Section IV. We conclude and discuss the implications and possible future applications of the method developed in this study in Section V. Additional detail of the parameters of VQE wavefunctions is presented in the appendix.

\section{Transverse field Ising Model}
\subsection{Model}
We consider a one-dimensional Ising model with transverse field. 
The Hamiltonian is given as

\begin{equation}
    H=-J\sum^{N}_{i=1}\hat{\sigma}^{z}_{i}\hat{\sigma}^{z}_{i+1}-\Gamma\sum_{i=1}^{N}\hat{\sigma}^{x}_{i},
\end{equation}
where $\hat{\sigma}^{\alpha}(\alpha=x,y,z)$ are the Pauli Matrices, which obey the commutation relation, $[\hat{\sigma}^{\alpha}_{i},\hat{\sigma}^{\beta}_{j}]=2 \iota \delta_{ij} \epsilon_{\alpha \beta \gamma}\hat{\sigma}^{\gamma}_{i}$, where $\iota$ is an imaginary number. $J$ is the coupling between the nearest neighbor spins, and is set to $1$ to serve as the energy scale of the problem. We only consider a ferromagnetic case with periodic boundary conditions in this study. 

$\hat{\sigma}^{z}$ has the eigenvalues as $\pm 1$ and their corresponding eigenvectors are symbolically denoted by 
\begin{equation}
    |\uparrow> = \begin{pmatrix} 1 \\
    0
    \end{pmatrix}
\end{equation}
and 
\begin{equation}
    |\downarrow> = \begin{pmatrix} 0 \\
    1
    \end{pmatrix}.
\end{equation}
%We can define the order parameter for ferromagnetism as $<\hat{\sigma}^{z}>$.

The model is solvable in the sense that the eigenenergy can be obtained exactly via the Jordan-Wigner transformation. The quantum critical point can also be determined exactly by mapping the model to anisotropic two-dimensional Ising model in a square lattice and employing the self-duality property of the model. The quantum critical point is at $\Gamma_{c}=J$ \cite{Kramers_Wannier_1941}. We will set $J$ equal to $1$ as the energy scale. Given the relative simplicity of the model and the value of the transverse field is exactly known at the quantum critical point, this provides a good test bed for the capability of a quantum classifier for identifying the phase transition of a quantum many-body system.

\subsection{Wavefucntion from VQE}
As our goal is to demonstrate the QCNN for identifying the wavefunction at different phases, the input is preferable to be represented in a quantum circuit. It is possible to cast the wavefunction in terms of a classical vector into quantum data. This is exactly what needs to be done using a quantum classifier for classical data, such as identifying classical images. 

We use the database provided by the  Tensorflow Quantum for the TFIM \cite{tf_quantum,tf}. There are in total 81 data points for $\Gamma=[0.2,1.8]$ in the spacing of $0.02$. The variational wavefunction is presented in the fig. \ref{fig:vqe}. Each qubit in the wavefunction represents a spin in the TFIM. The first layer of the quantum circuit is composed of a Hadamard transform by adding a Hadamard gate to each qubit. Then, $N/2$ layers of gates are acted on the quantum circuit. Each layer contains two sublayers. The first sublayer is composed of $N$ ZZ Ising coupling gates, which are then acting on each pair of nearest neighbor spins represented by the corresponding qubits. The rotation angle is fixed for each ZZ Ising coupling gate within the same sublayer. The second sublayer is composed of $N$ Pauli X gates, each of them acting on each qubit with a fixed rotation angle within the sublayer. Therefore, in total, the quantum circuit contains $N/2 \times (N+N)=N$ variational parameters. 

Only $81$ data points are available in the Tensorflow Quantum database, which is a rather small number for the application of ML \cite{tf_quantum,Cirq}. We generate additional data points for finer grids of $\Gamma$ points, so that more data is available for the training and testing of the QCNN. We did not perform the optimization as that in the standard VQE method to obtain the new data. We take advantage of the fact that all the variational parameters vary smoothly as a function of the transverse field $\Gamma$ in the TFIM. (See the appendix for the variational parameters as a function of $\Gamma$) The additional data points are simply generated by linear interpolation of the variational parameters with respect to $\Gamma$.

\section{QCNN}
By generating a larger set of data points, we are then able to train a QCNN. The input for our QCNN model is presented in the form of a quantum circuit. This quantum circuit is the representation of our wavefunction from the VQE solver for the TFIM. 

The QCNN is another quantum circuit constructed with an alternating series of convolutional and pooling layers, until the number of pooling layers dwindles to a single qubit. The idea is that the quantum circuit can reap important information from the input quantum data.
%The QCNN quantum circuit also depends on some tunable parameters, which are adjusted during training. 

The goal here is to prepend our data points to a QCNN model and train it to correctly identify the `phase' of each wavefunction. This is performed in a supervised environment, as we already have the correct phase identification for each data point given in the database. For a given transverse field, we have either a ferromagnetic or paramagnetic phase. In the range of $\Gamma=[0.2,1.8]$,  the system is in a ferromagnetic phase below the value of $1$, and a paramagnetic phase above the value of $1$.

The pooling unit we employed for the present study, which pools two qubits into one qubit, is shown in the fig. \ref{fig:pool_layer}. It consists of local rotation gates on each qubit, and then a controlled X-gate, and the inverse rotation on the controlled bit. Thus, in total there are 6 parameters in each pooling unit which pools two qubits into one qubit.

\begin{widetext}

\begin{figure}[htpb]
    \centering
        \includegraphics[height=3.0cm]{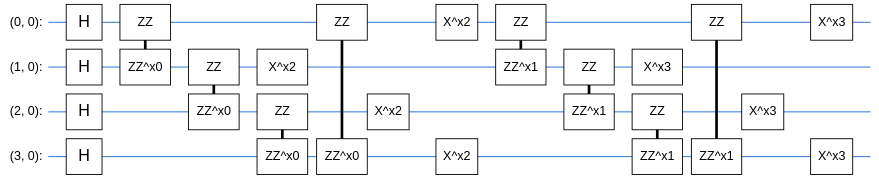}
    \caption{Variational wavefunction for the TFIM in the Tensorflow Quantum database, for $N=4$ \cite{tf_quantum}. The notation of the quantum gates is adopted from the Cirq \cite{Cirq}. H is the Hadamard gate, ZZ is the Ising coupling gate with the rotation angle given by the associated number, X is the Pauli X gate with the rotation angle associated with the number. $x0,x1,x2,x3$ are the four parameters for the wavefunction. }
    \label{fig:vqe}
\end{figure}

\begin{figure}[htpb]
    \centering
        \includegraphics[height=1.8cm]{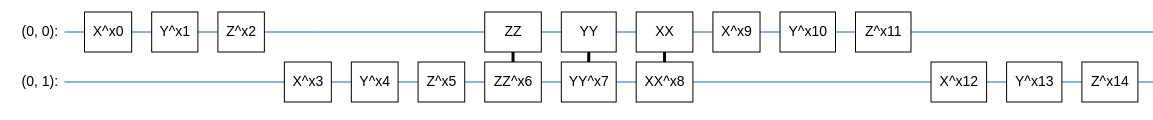}
    \caption{Convolution between two qubits. A convolutional layer of $N$ qubits is composed of $N/2$ convolutional units acting on the qubits which represent pairs of nearest neighbor spins. $x0, x1, ..., x14$ are the parameters.}
    \label{fig:conv_layer}
\end{figure}

\begin{figure}[htpb]
    \centering
        \includegraphics[height=1.8cm]{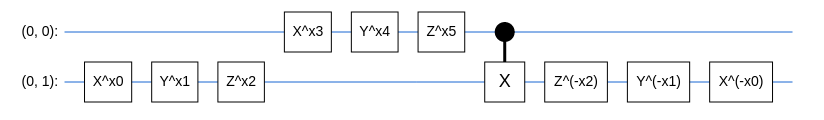}
    \caption{Pooling between two qubits. A pooling layer from $N$ qubits to $N/2$ qubit is composed of $N/2$ of pooling units. $x0, x1, ...,x5$ are the parameters. }
    \label{fig:pool_layer}
\end{figure}

\end{widetext}

The convolutional layer is composed of multiple two qubit convolution units. A convolution unit for two qubits is shown in the fig. \ref{fig:conv_layer}. It consists of local rotation gates on each qubit sandwiched between the Ising coupling gates between the two qubits. Therefore, in total, there are $15$ parameters for each convolutional unit between two qubits.

The code for the QCNN is written in Cirq and using the Tensorflow Quantum package for training \cite{Cirq,tf_quantum}.

\begin{figure}[t]
    \centering
        \includegraphics[height=2.75cm, trim = 0.5cm 5cm 0.5cm 2cm]{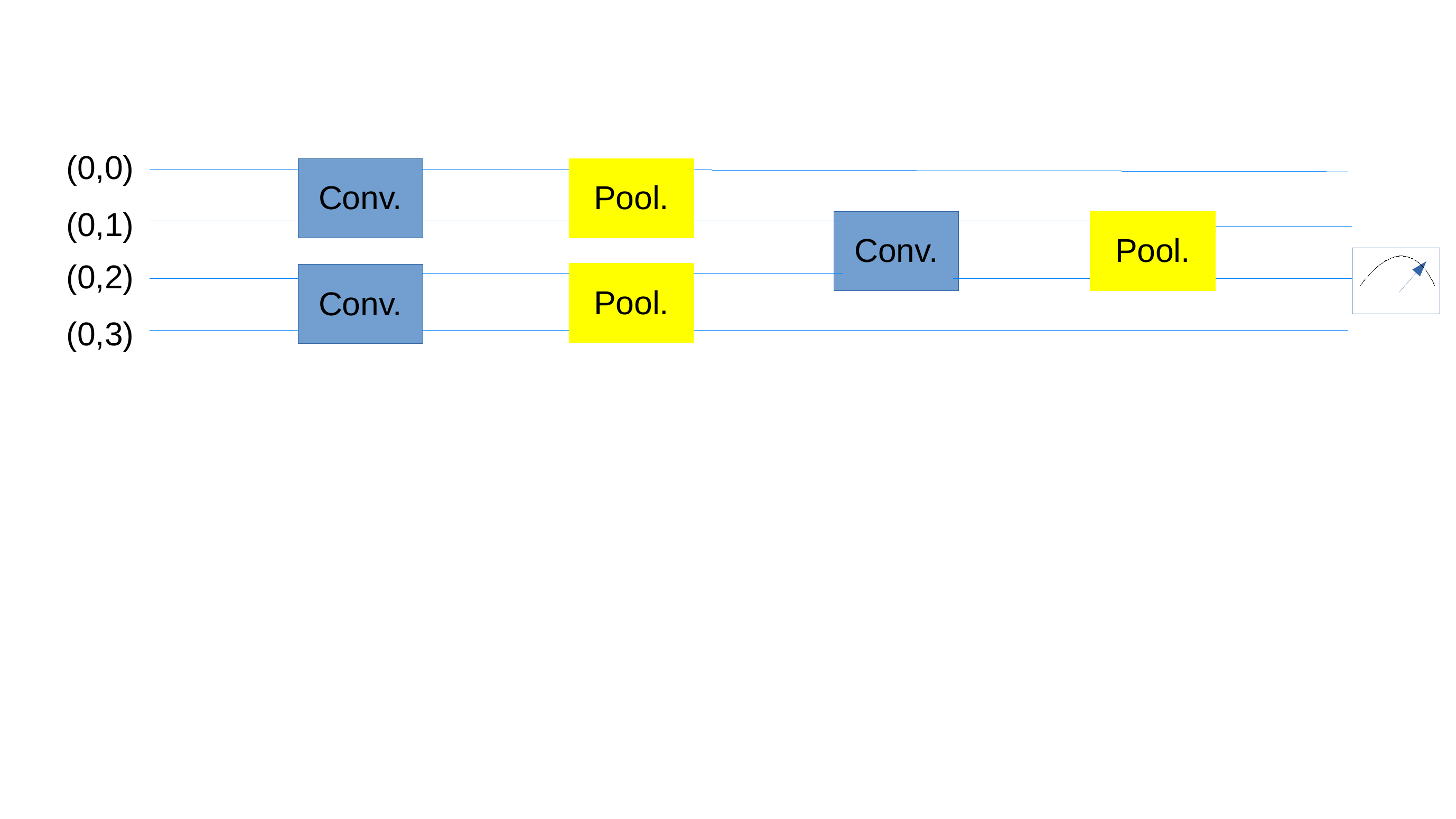}
    \caption{An example of the entire QCNN circuit for 4 qubits, two convolutional units act on the pairs of nearest neighbor qubits. A pooling unit which reduces the 4 input qubits into 2 output qubits. Another layer of convolutional units on the nearest neighbor pair of qubits, and the pair of qubits is then fed into another pooling unit with one output qubit. The final qubit is measured. Note that the parameters in a convolutional units can be different from the other, so do the pooling untis.}
    \label{fig:qcnn}
\end{figure}

%\begin{widetext}
%\begin{figure}
%    \centering
%        \includegraphics[width=18.cm]{QCNN/1.png}
%        \includegraphics[width=18.cm]{QCNN/2.png}
%        \includegraphics[width=18.cm]{QCNN/3.png}
%        \includegraphics[width=18.cm]{QCNN/4.png}
%        \includegraphics[width=18.cm]{QCNN/5.png}
%        \includegraphics[width=18.cm]{QCNN/6.png}
%        \includegraphics[width=18.cm]{QCNN/7.png}
%        \includegraphics[width=18.cm]{QCNN/8.png}
%        \includegraphics[width=18.cm]{QCNN/9.png}
%        \includegraphics[width=18.cm]{QCNN/10.png}
%    \caption{QCNN for classifying the VQE wavefunction for %the TFIM.   }
%    \label{fig:2}
%\end{figure}
%\end{widetext}

% --------------------
\section{Results}
% --------------------
We trained the QCNN by using two different sets of training data. For the first one, the training of the QCNN is proceeded by randomly picking $80\%$ of the wavefunctions with labels to designate their corresponding phases as the training data set. For the second one, only the data points corresponding to small and large transverse field are used for training. We define the label for the ferromagnetic phase as $-1$, and the paramagnetic phase as $+1$. For benchmarking the accuracy of predictions, we cast the output to $-1$ if the measurement of the output from the QCNN is smaller than $0$ and similarly we cast the output to $+1$ if the measurement is larger than $0$.

\subsection{Training QCNN with data for randomly picked data for $0.2 \leq \Gamma \leq 1.8$}

The accuracy of the trained QCNN is benchmarked against the remaining $20\%$ of the available samples. We show the loss and accuracy during each iteration of the training processes for three system sizes $N=4,8,$ and $12$.

\begin{figure}[htpb]
    \centering
        \includegraphics[width=7cm]{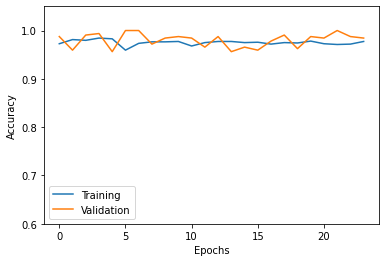}
        \includegraphics[width=7cm]{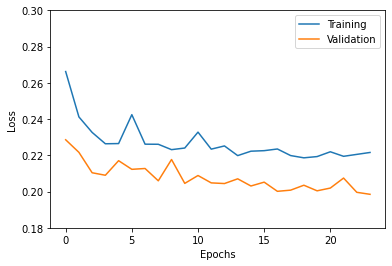}
    \caption{Accuracy and loss for the training and validation data sets the QCNN for $N=4$ as a function of the number of epochs.}
    \label{fig:3}
\end{figure}

\begin{figure}[htpb]
    \centering
        \includegraphics[width=7cm]{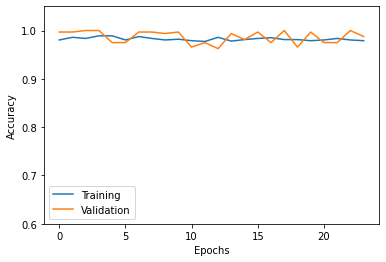}
        \includegraphics[width=7cm]{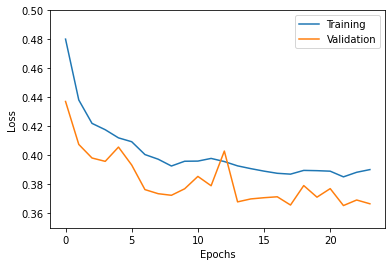}
    \caption{Accuracy and loss for the training and validation data sets the QCNN for $N=8$ as a function of the number of epochs.}
    \label{fig:4}
\end{figure}

\begin{figure}[htpb]
    \centering
        \includegraphics[width=7cm]{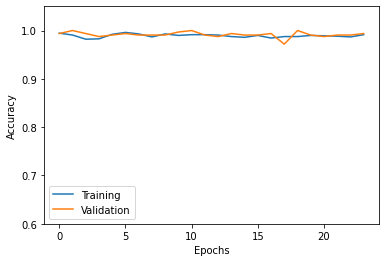}
        \includegraphics[width=7cm]{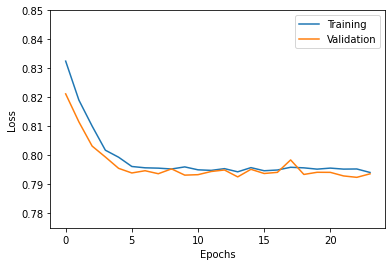}
    \caption{Accuracy and loss for the training and validation data sets the QCNN for $N=12$ as a function of the number of epochs.}
    \label{fig:5}
\end{figure}

By training our QCNN with randomly chosen wavefunctions, we allow our network to study samples across the entire dataset, and the QCNN will become familiar with values of $\Gamma$ both close and far from the quantum critical point. In doing so, we expect the QCNN will have a familiarity with the wavefunctions, giving us valid results for the trained data predictions. When using a system size of $N = 4$ (see fig. \ref{fig:3}), this is exactly what we observed. The accuracy for both training and testing data is consistently high, and we observe minimal fluctuations in the accuracy from start to finish. 
We can conclude that this method of randomized data allowed the QCNN to adjust to the variations in wavefunctions during training, and then apply this to our testing data . We observe very similar results for the accuracy following with system sizes $N = 8$ and $12$ (see figs. \ref{fig:4} and \ref{fig:5}). Increasing the system size has no significant impact on the results for accuracy. The ability to predict the `phase' of wavefunctions proved to be a task our QCNN is capable of executing.


We can observe what happens as the QCNN undergoes the training phase when studying the results for the network’s loss. As stated, the network is fully trained once the loss converges. This means that once the loss stops decreasing, the QCNN is fully trained and at its maximum potential. The results for the loss of system $N = 4$ show how the loss decreases at a high rate during the initial training iterations, and then it begins to flatten as the QCNN reaches its fullest potential. We begin to notice a pattern by observing larger system sizes. As we include more input variables, the initial and final loss grows. %This is to be expected in a machine learning environment, as more input variables allows more room for the loss to grow. 
Regardless of the increased value of loss, we can still observe how the system is trained during each iteration. All the system sizes that are tested gave results that allowed a visualization of the QCNN’s training phase, and they showed that the network is capable of enhancing its ability to give us correct predictions for the `phase' corresponding to the wavefunctions.

\subsection{Training with data for $0.5<\Gamma$ and $\Gamma>1.5$ }

Following the above results, we decide testing the QCNN with the chosen training and testing data. For training, we use wavefunctions that have values of $\Gamma$ below $0.5$ and above $1.5$. Testing data is in the range of $[0.5, 1.5]$. This would allow us to observe how the QCNN behaved when classifying data near the known quantum critical point ($\Gamma_c=1$) after being trained with data far from the quantum critical point . For each system size, we observe one hundred percent accuracy during the training phase.
This is perhaps not surprising given that the two sets of data for very large and very small $\Gamma$ are far apart from each others.

\begin{figure}
    \centering
    \includegraphics[width=7cm]{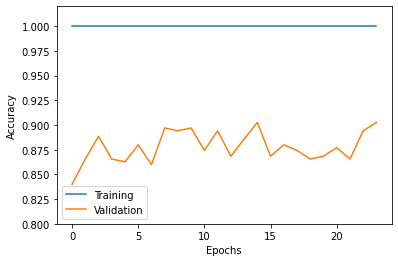}
    \includegraphics[width=7cm]{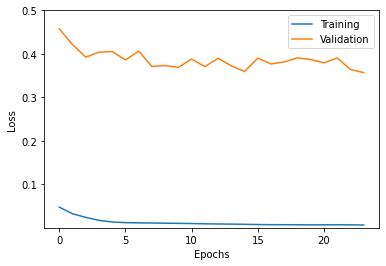}
    \caption{Accuracy and loss for the training and validation data sets the QCNN for $N=4$ with training data $\Gamma<0.5$ and $\Gamma>1.5$ as a function of the number of epochs.}
    \label{fig:6}
\end{figure}

\begin{figure}
    \centering
    \includegraphics[width=7cm]{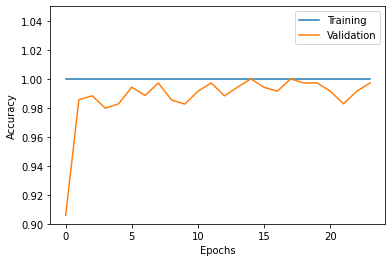}
    \includegraphics[width=7cm]{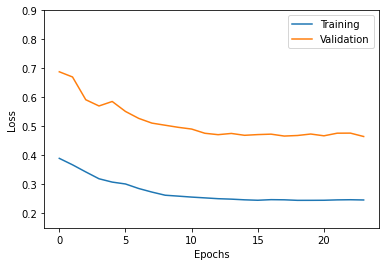}
    \caption{Accuracy and loss for the training and validation data sets the QCNN for $N=8$ with training data $\Gamma<0.5$ and $\Gamma>1.5$ as a function of the number of epochs.}
    \label{fig:7}
\end{figure}

\begin{figure}
    \centering
    \includegraphics[width=7cm]{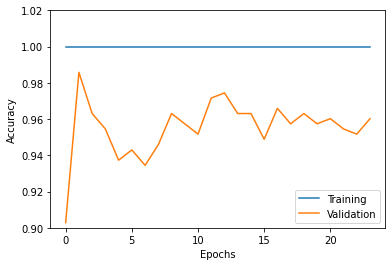}
    \includegraphics[width=7cm]{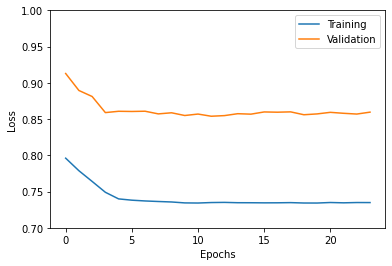}
    \caption{Accuracy and loss for the training and validation data sets the QCNN for $N=12$ with training data $\Gamma<0.5$ and $\Gamma>1.5$ as a function of the number of epochs.}
    \label{fig:8}
\end{figure}

We plot the results for $N=4, 8,$ and $12$ in the figs. \ref{fig:6}, \ref{fig:7}, and \ref{fig:8} respectively. 
The testing data was consistently lower than when using a randomized data set, but it showed a sharp increase during its initial iterations. The high accuracy in training shows that the QCNN is capable of classifying data points far from the quantum critical point. As we approach the quantum critical point, the network has more difficulty with predictions. The inability to familiarize itself with data points in this range hindered the accuracy for testing data. The sharp increase shows that after encountering a few data points near quantum critical, the network can learn to improve its capability in our range of $[0.5, 1.5]$. The loss shows small changes during the training phase compared to using randomized data. While we still observe a decrease, this decrease is minimal. The QCNN does not have much room for improvement when the testing data is isolated from the quantum critical point.

The result of using only large and small values of $\Gamma$ for training the QCNN is significant for the prospect of using QCNN to detect the quantum critical point. A rather well-studied scheme for using supervised classical ML to detect phase transitions is to train a supervised classical ML, such as CNN, for the control parameters of a system, for example, temperature in the thermal transition and external parameter in the quantum phase transition, away from the putative phase transition or critical point \cite{Carrasquilla_etal_2017}. The results here demonstrate that the QCNN can also be trained by using data away from the putative quantum critical point with good accuracy by predicting the phases around the critical point. 

%For this purpose, we present the results of predicting the corresponding phase of the wavefunctions generated for $0.5 \geq \Gamma \leq 1.5$ by the QCNN trained with wavefunctions generated for $0.5 < \Gamma$ and $\Gamma > 1.5$.


\subsection{Predicted labels as a function of $\Gamma$ }

With the intent of locating the quantum critical point, we can plot the predicted values from our QCNN to find a pattern in the outputs. In this experiment, we aim to find the value of $\Gamma$ where the QCNN predicted labels switch between the two phases, ferromagnetic and paramagnetic. This would mark a good approximation to the quantum critical point. The quantum phase transition occurs at $\Gamma=1$ for the TFIM, and we aim to observe this with the results given by the QCNN. In order to find if the network is capable of this, a plot will need to be made describing how the predicted labels changes as $\Gamma$ changes. We can see if the phase transition happens at or near $\Gamma=1$, allowing us to conclude the validity of the QCNN.

\begin{figure}[h]
    \centering
    \includegraphics[width=7cm]{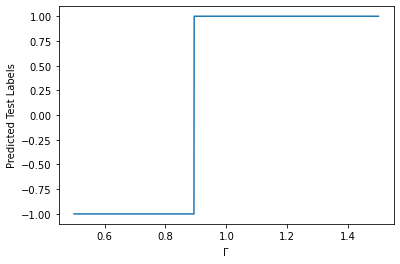}
    \caption{Predicted test labels vs $\Gamma$ for $N=4$. The predicted labels jump at $\Gamma=0.89$.}
    \label{fig:9}
\end{figure}

\begin{figure}[h]
    \centering
    \includegraphics[width=7cm]{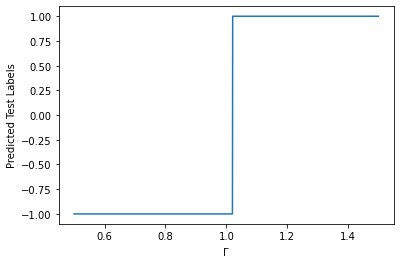}
    \caption{Predicted test labels vs $\Gamma$ for $N=8$.  The predicted labels jump at $\Gamma=1.02$.}
    \label{fig:10}
\end{figure}

\begin{figure}[h]
    \centering
    \includegraphics[width=7cm]{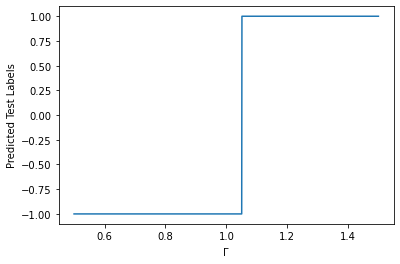}
    \caption{Predicted test labels vs $\Gamma$ for $N=12$. The predicted labels jump at $\Gamma=1.05$.}
    \label{fig:11}
\end{figure}

Observing these transitions for systems $N=4,8,12$ in figs. \ref{fig:9}, \ref{fig:10}, \ref{fig:11}, we notice that each trial holds a transition near $\Gamma=1$. This allows us to deduce the value of $\Gamma$ that holds a quantum critical point. The QCNN predicts a ferromagnetic phase for low values of $\Gamma$, and it predicts a paramagnetic phase for high values of $\Gamma$. We can see the network output ferromagnetic prediction for $\Gamma<1$. As $\Gamma$ approaches $1$, we expect to find a quantum critical point. This can be identified by a sudden jump to a different output value signifying a paramagnetic phase. In figs. \ref{fig:9}, \ref{fig:10}, \ref{fig:11}, this jump can be seen near $\Gamma=1$. We observe a series of ferromagnetic predictions, and then a series of paramagnetic predictions. The value of $\Gamma$ where the predicted phase change occurs is our approximation to the quantum critical point. As our predicted quantum critical point consistently lies near the true value, we can conclude the network is capable of being used to identify quantum critical points.

Thus, the technology developed for using classical supervised ML can be used here for detecting quantum critical points \cite{Carrasquilla_etal_2017}, the major modification is to replace the classical supervised ML, such as the classical CNN, by the QCNN.


 % --------------------
\section{Discussion and Conclusions}
% --------------------

From the given results, we can conclude that the QCNN model is capable of detecting whether a TFIM wavefunction is ferromagnetic or paramagnetic.  We can train a QCNN model to detect the represented phase of the wavefunction by implementing a dataset of varying external parameters. Once the training phase has completed, we will have a fully trained network that provides an efficient method for phase detection using the TFIM.

The study of quantum phase transitions has been one of the major topics in condensed matter physics for multiple reasons. First, devising an effective theory for describing the quantum critical point is often not a simple task. Various approaches have been brought forth to understand the quantum critical point from a straightforward renormalization group from the upper critical dimension to the more exotic gauge gravity duality approaches \cite{Hertz_1976,Sachdev_2011}. Second, the quantum critical point is believed to be responsible for many exotic behaviors in strongly correlated systems, particular examples including the high temperature superconductivity in cuprates \cite{vidhyadhiraja_etal_2009}, and non-Fermi liquids \cite{Varma_etal_2002}. The study of the properties or even detecting the existence of a quantum critical point has been a major topic in computational condensed matter physics for models from two to infinite dimensions \cite{vidhyadhiraja_etal_2009,Kellar_Tam_2020,Terletska_etal_2011,Fuchs_etal_2011}.

The encouraging results of using QCNN to identify the ferromagnetic phase and the paramagnetic phase of the TFIM from the wavefunction obtained by VQE presents an opportunity which is rather different from other conventional computational approaches, such as quantum Monte Carlo and other diagonalization based methods. 

Whether such an approach is applicable to more interesting and complicated models, such as Hubbard model beyond one dimension, is a crucial question to be addressed by future studies. The two key questions are 1. Whether the VQE can provide a sufficiently accurate wavefunction for strongly correlated problems \cite{Kardashin_etal_2021}. 2. Whether the wavefunction has sufficient features which can be identified by some form of quantum classifier. This is more acute due to the limited system size of the models which can be simulated in the NISQ machines, as ground state energy alone or even order parameters may not be very helpful for identifying phase transition for small system sizes.

The nature of the Hamiltonian of chemistry problems is in general rather different from that of strongly correlated systems. The Hamiltonian for molecules are often diagonally dominated in the sense that the off-diagonal matrix elements are small compared to the diagonal ones. On the other hand, models for strongly correlated systems often have comparable off-diagonal and diagonal matrix elements. That is perhaps the reason some of the highly successful methods in computational chemistry, such as conventional coupled cluster theory, have seen limited success for strongly correlated systems. Therefore, whether the present approach can be transplanted to strongly correlated systems, which are relevant to a plethora of exotic properties of materials, remains to be examined. 

\section{Acknowledgment}
This manuscript is based upon the work supported by NSF DMR-1728457. This work used the high performance computational resources provided by the Louisiana Optical Network Initiative (http://www.loni.org), and HPC@LSU computing. JM and KMT are partially supported by the U.S. Department of Energy, Office of Science, Office of Basic Energy Sciences under Award Number DE-SC0017861. This material is based upon work supported by the National Science Foundation under the award OAC-1852454 with additional support from the Center for Computation \& Technology at Louisiana State University.
%\lipsum[14] % Dummy text. Erase before write

% %%%%%%%%%%%%%%%%%%%%%%%%%%%%%%%%%%%%%%%%%%%%%%%%%%%%%%%%%%
% %%%%%%%%%%%%%%%%%%%%%%%%%%%%%%%%%%%%%%%%%%%%%%%%%%%%%%%%%%
% REFERENCES SECTION
% %%%%%%%%%%%%%%%%%%%%%%%%%%%%%%%%%%%%%%%%%%%%%%%%%%%%%%%%%%
% %%%%%%%%%%%%%%%%%%%%%%%%%%%%%%%%%%%%%%%%%%%%%%%%%%%%%%%%%%
\medskip
\appendix*

\section{Variational parameters}
\begin{figure}
    \centering
        \includegraphics[width=7.5cm]{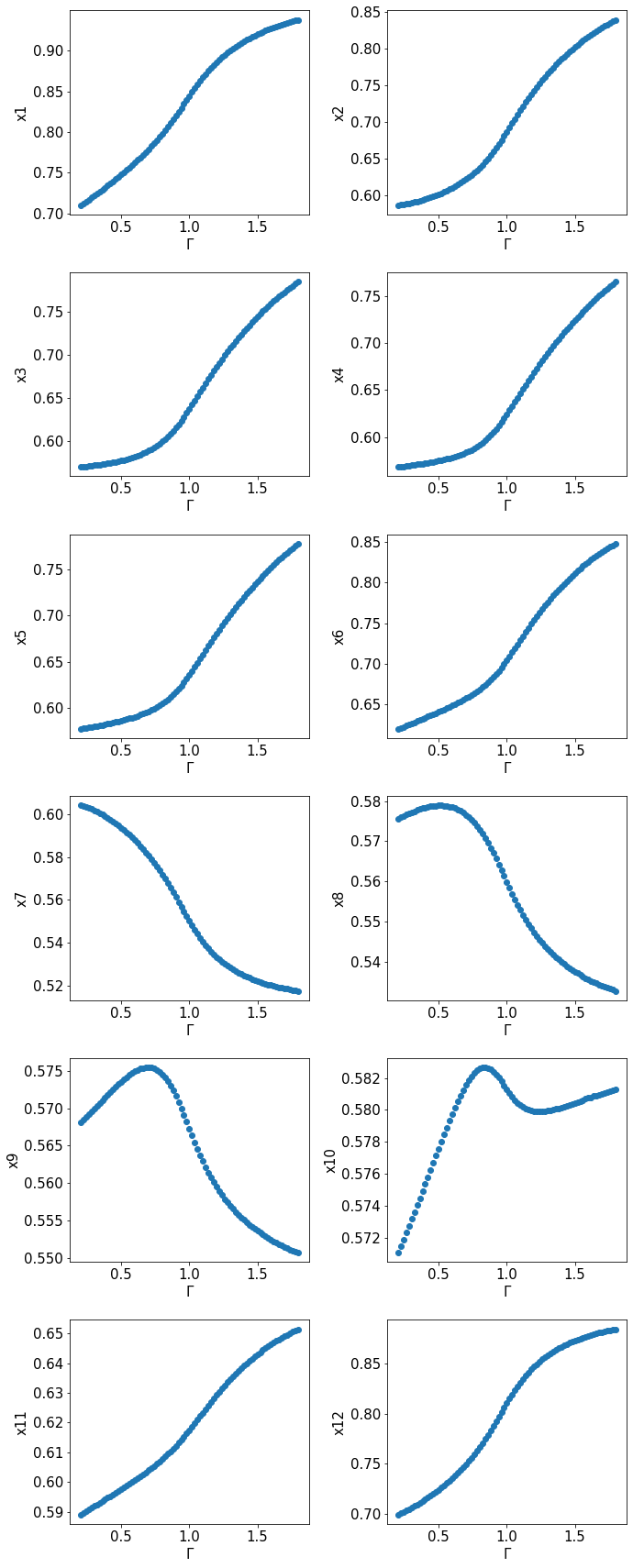}
        \caption{Plot of the variational parameters ($x1,x2,...x12$) of the VQE wavefunctions in the TensorFlow Quantum database for $N=12$ transverse field Ising model as a function of the transverse field ($\Gamma$). All the variational parameters for $N=12$, all of them change smoothly with respect to $\Gamma$. We generate additional wavefunctions by simple linear interpolation of these parameters with respect to $\Gamma$. The parameters for $N=4$ and $8$ are also smooth with respect to $\Gamma$ (not shown here).}
    \label{fig:params}
\end{figure}

The VQE wavefunction of the TFIM is extracted from the database in the Tensorflow Quantum \cite{tf_quantum}. We only consider the cases of $N=4, 8,$ and $12$ for periodic boundary conditions. For a fixed system size, $N$, the structure of the quantum circuit which represents the wavefunction is independent of the value of the transverse field strength, $\Gamma$. The database from the Tensorflow Quantum contains 81 data points for $\Gamma = 0.2, 0.22, \cdots, 1.8$. For the purpose of training a supervised machine learning classifier, it is desirable to have a much larger data set with a much finer grid of $\Gamma$ points. We utilize the fact that all the parameters in the variational wavefunction are smooth with respect to $\Gamma$ to generate additional wavefunctions for finer grids of $\Gamma$ by simple linear interpolation. We plot all the $12$ parameters from the Tensorflow Quantum database for case of $N=12$ in fig. \ref{fig:params}. All the variational parameters not shown here for $N=4$ and $8$ are also smooth with respect to $\Gamma$.

\newpage 
\bibliography{references.bib} 

\newpage

\end{document}